\documentstyle[aps, preprint,eqsecnum, graphicx, verbatim]{revtex}
\tightenlines
\begin{document}
%\draft
\title{Chiral stabilization of the renormalization group 
for flavor and color anisotropic 
current interactions}
\author{
Andr\'e  LeClair }
\bigskip
\address{ Newman Laboratory, Cornell University, Ithaca, NY
14853.}
\medskip
\date{May  2001}
\maketitle

\def\betaf{{$\beta$eta~}}
\def\bar{\overline}
\def\Jbar{\bar{J}}

\begin{abstract}

We propose an all-orders \betaf function for current-current
interactions in 2d  with flavor anisotropy.  
When the number of left-moving and right-moving flavors are
unequal, the \betaf function has a non-trivial fixed point
at finite values of the couplings.   We also extend the computation
to simple cases with both flavor and color anisotropy.

\end{abstract}
\vskip 0.2cm
%\narrowtext
%               DEFINITIONS FOR TEX
%
%%%%%%%%%%%%%%%%%%%%%%%%%%%%%%%%%%%%%%%%%%%%%%%%%%%%%%%%%%%%%%%
%
%
%\def\e{\'e}
%\def\ee{\`e}
%%%%%%%%%%%%%%%%%%%DEFINITIONS%%%%%%%%%%%%%%%%%%%%%%%%%%%%%%%%%
%
\def\Jb{\bar{J}}
\def\dx{\frac{d^2x}{2\pi}}
\def\oti{{\otimes}}
\def\bra#1{{\langle #1 |  }}
\def\lb{ \left[ }
\def\rb{ \right]  }
\def\tilde{\widetilde}
\def\hat{\widehat}
\def\*{\star}
\def\[{\left[}
\def\]{\right]}
\def\({\left(}          \def\BL{\Bigr(}
\def\){\right)}         \def\BR{\Bigr)}
        \def\BBL{\lb}
        \def\BBR{\rb}
%
%%%%%%%%%%%%%%%%%%%%%%%%%%%%%%%%%%%%%%%%%%%%%%%%%%%%%%%%%%%%%%%
%
\def\zb{{\bar{z} }}
\def\zbar{{\bar{z} }}
\def\frac#1#2{{#1 \over #2}}
\def\inv#1{{1 \over #1}}
\def\half{{1 \over 2}}
\def\d{\partial}
\def\der#1{{\partial \over \partial #1}}
\def\dd#1#2{{\partial #1 \over \partial #2}}
\def\vev#1{\langle #1 \rangle}
\def\ket#1{ | #1 \rangle}
\def\rvac{\hbox{$\vert 0\rangle$}}
\def\lvac{\hbox{$\langle 0 \vert $}}
\def\2pi{\hbox{$2\pi i$}}
\def\e#1{{\rm e}^{^{\textstyle #1}}}
\def\grad#1{\,\nabla\!_{{#1}}\,}
\def\dsl{\raise.15ex\hbox{/}\kern-.57em\partial}
\def\Dsl{\,\raise.15ex\hbox{/}\mkern-.13.5mu D}
%
%%%%%%%%%%%%%%%%%%%%GREEK LETTERS%%%%%%%%%%%%%%%%%%%%%%%%%%%%%%
%
\def\th{\theta}         \def\Th{\Theta}
\def\ga{\gamma}         \def\Ga{\Gamma}
\def\be{\beta}
\def\al{\alpha}
\def\ep{\epsilon}
\def\vep{\varepsilon}
\def\la{\lambda}        \def\La{\Lambda}
\def\de{\delta}         \def\De{\Delta}
\def\om{\omega}         \def\Om{\Omega}
\def\sig{\sigma}        \def\Sig{\Sigma}
\def\vphi{\varphi}
%
%%%%%%%%%%%%%%%%%%%CALIGRAPHIC LETTERS%%%%%%%%%%%%%%%%%%%%%%%%%
%
\def\CA{{\cal A}}       \def\CB{{\cal B}}       \def\CC{{\cal C}}
\def\CD{{\cal D}}       \def\CE{{\cal E}}       \def\CF{{\cal F}}
\def\CG{{\cal G}}       \def\CH{{\cal H}}       \def\CI{{\cal J}}
\def\CJ{{\cal J}}       \def\CK{{\cal K}}       \def\CL{{\cal L}}
\def\CM{{\cal M}}       \def\CN{{\cal N}}       \def\CO{{\cal O}}
\def\CP{{\cal P}}       \def\CQ{{\cal Q}}       \def\CR{{\cal R}}
\def\CS{{\cal S}}       \def\CT{{\cal T}}       \def\CU{{\cal U}}
\def\CV{{\cal V}}       \def\CW{{\cal W}}       \def\CX{{\cal X}}
\def\CY{{\cal Y}}       \def\CZ{{\cal Z}}

\def\rvac{\hbox{$\vert 0\rangle$}}
\def\lvac{\hbox{$\langle 0 \vert $}}
\def\comm#1#2{ \BBL\ #1\ ,\ #2 \BBR }
\def\2pi{\hbox{$2\pi i$}}
\def\e#1{{\rm e}^{^{\textstyle #1}}}
\def\grad#1{\,\nabla\!_{{#1}}\,}
\def\dsl{\raise.15ex\hbox{/}\kern-.57em\partial}
\def\Dsl{\,\raise.15ex\hbox{/}\mkern-.13.5mu D}
%
%%%%%%%%%%%%%%%%%%%%GREEK LETTERS%%%%%%%%%%%%%%%%%%%%%%%%%%%%%%
%
%%%%%%%%%%%%%%% MATH CHARACTERS %%%%%%%%%%%%%%%%%%%%%%%%%%%%
%
\font\numbers=cmss12
%\font\numbers=cmu10 scaled\magstep1
\font\upright=cmu10 scaled\magstep1
\def\stroke{\vrule height8pt width0.4pt depth-0.1pt}
\def\topfleck{\vrule height8pt width0.5pt depth-5.9pt}
\def\botfleck{\vrule height2pt width0.5pt depth0.1pt}
\def\Zmath{\vcenter{\hbox{\numbers\rlap{\rlap{Z}\kern
0.8pt\topfleck}\kern 2.2pt
                   \rlap Z\kern 6pt\botfleck\kern 1pt}}}
\def\Qmath{\vcenter{\hbox{\upright\rlap{\rlap{Q}\kern
                   3.8pt\stroke}\phantom{Q}}}}
\def\Nmath{\vcenter{\hbox{\upright\rlap{I}\kern 1.7pt N}}}
\def\Cmath{\vcenter{\hbox{\upright\rlap{\rlap{C}\kern
                   3.8pt\stroke}\phantom{C}}}}
\def\Rmath{\vcenter{\hbox{\upright\rlap{I}\kern 1.7pt R}}}
\def\Z{\ifmmode\Zmath\else$\Zmath$\fi}
\def\Q{\ifmmode\Qmath\else$\Qmath$\fi}
\def\N{\ifmmode\Nmath\else$\Nmath$\fi}
\def\C{\ifmmode\Cmath\else$\Cmath$\fi}
\def\R{\ifmmode\Rmath\else$\Rmath$\fi}
%%%%%%%%%%%%%%%%%%%%%%%%%%%%%%%%%%%%%%%%%%%%%%%%%%%%%%%%%%%%%%%%%
 %%%%%%%%%%%%%%%%%% END OF DEFINITIONS %%%%%%%%%%%%%%%%%%%%%%
 %%%%%%%%%%%%%%%%%%%%%%%%%%%%%%%%%%%%%%%%%%%%%%%%%

%\section{Introduction}

%Bla..Bla Bla\cite{BPZ}. 

%\begin{equation}
%\label{hamilo}
%H=H^{atom}+H^{field}+H^{int}
%\end{equation}
%with
%\begin{eqnarray}
%H^{field}&=&\frac{1}{2} \int_{-\infty}^\infty
%dx [(\partial_t \phi)^2+(\partial_x\phi)^2] ,
%\\ \nonumber
%H^{atom}&=&\frac{ \omega_0}{2} \sigma_3 ,
%{}~~~~~H^{int}=\frac{\beta}{2} \partial_t \phi (x_0) \left( \sigma_++
%\sigma_- \right),
%\label{hamili}
%\end{eqnarray}

%The hamiltonian (\ref{hamili}) 

%\vbox{
%\epsfysize=8cm
%\epsfxsize=8cm
%\epsffile{classlin.eps}
%\begin{figure}
%\caption[]{\label{fig1} Accuracy comparison for various
%form factor contributions for $g=1/5$.}
%\end{figure}}

%\begin{references}
%\bibitem{BPZ}  Belavin, Polyakov and Zamolodchickov.
%\end{references}

%\end{document}

\def\beq{\begin{equation}}
\def\eeq{\end{equation}}
\def\cg{{\cal G}}
\def\ch{{\cal H}}
\def\barray{\begin{eqnarray}}
\def\earray{\end{eqnarray}}

\section{Introduction}

In \cite{GLM}  an all-orders  \betaf function was proposed
for general Lie group $\CG$  anisotropic current-current interactions
in two dimensions.  Non-perturbative  aspects of the resulting
RG flows were studied in ref. \cite{BLdual} a strong-weak coupling
duality.   
    These \betaf functions generally have
no non-trivial zeros, i.e. fixed points, at intermediate
values of the couplings between $0$ and $\infty$.  Rather, 
interesting fixed points arise if one is attracted 
under renormalization group (RG)  flow 
to sub-manifolds corresponding to a sub-group $\CH$ of $\CG$,
the fixed point being the current-algebra coset $\CG/\CH$\cite{Lec}
as $g\to \infty$.  
  
Referring to $\CG$ anisotropy as color anisotropy,  in this
paper we consider anisotropy in the number of copies, or flavors,
of the current algebra.   The simplest form of flavor anisotropy
corresponds to unequal numbers of left verses right-moving 
flavors $N_L$ and $N_R$.  Unequal numbers of chiral flavors
is equivalent to having unequal levels $k_L, k_R$ for
the current algebras in the left verses right sector. 
Based on the work of Polyakov-Wiegmann\cite{PW},
one expects a non-trivial fixed point of the \betaf function 
as $N_L , N_R \to \infty$ with  $N_R - N_L $ fixed which
  corresponds 
to the WZW model at level $k=N_R-N_L$.     
The generalization of this fixed point when $N_L , N_R$ are not
$\infty$ was proposed in \cite{Andrei}\cite{Andrei2}, where the
phenomenon 
was referred to as chiral stabilization.  
In ref. \cite{Andrei} general 
arguments based on the preservation of the difference 
between  left-right Virasoro
central charges $c_R - c_L$ in the RG flow 
led to a precise identification of
the infra-red (IR) fixed point as a chirally
asymmetric coset;  this identification was further
supported by a thermodynamic Bethe ansatz analysis 
for the integrable cases.   Evidence for these fixed points
based on the \betaf function were not given in \cite{Andrei} 
presumably because the fixed point cannot be seen at one-loop.             

We first  extend the computation of the \betaf function
in \cite{GLM} to models with
flavor anisotropy but isotropic in color.   For equal numbers
of left and right-moving flavors we do not find any non-trivial
fixed points for the examples we have looked at. 
For unequal numbers of left-right flavors, 
we find the fixed points  predicted in \cite{PW}\cite{Andrei}.   
The anomalous dimension of the perturbing operator in the infra-red
(IR)  is 
a strong test of the proposed \betaf function.   
We find that the \betaf function gives the expected result except 
for the well-known shift of the level
$k$ by the dual Coxeter number in the affine-Sugawara 
construction\cite{Olive}.   We remark on the possible sources
of this discrepancy but do not resolve it in this paper.               

We also compute the \betaf function for color anisotropic
models where the only flavor anisotropy corresponds to unequal
numbers of left verses right movers, or equivalently 
unequal levels $k_L ,k_R$.
This corresponds
to the models in \cite{GLM} with the difference that the left
and right moving levels of the current algebra are unequal.   
Here also we find that this chiral anisotropy stabilizes the 
RG flow in that it leads to fixed points at intermediate values
of the coupling that are generally on color isotropic manifolds.   

In the next section we propose an exact formula for the most
general form of flavor anisotropy that preserves the color symmetry.  
In section III  we consider the simplest possible chirally
asymmetric case of one
left-moving flavor and two right moving flavors.  
The most generic anisotropic model  flows to a fixed point 
that is isotropic in the flavor couplings, the only remaining
anisotropy being in the number of flavors.  The model
then flows to the fixed points of the kind described in
\cite{PW}\cite{Andrei}.    In section IV 
we extend the computation to the  special case of both flavor and
color anisotropy mentioned above, namely
unequal levels $k_L, k_R$.     Here the  chirality stabilizes
the flow in two ways, namely the flow to the isotropic line is
stabilized whereas for $N_L = N_R$ it is unstable\cite{BLdual},
and also once on the isotropic line the flow off to infinity
is prevented by the fixed point.

\section{General flavor anisotropy}

Let us denote by $\CG_k$ the $\CG$ current algebra of level $k$
and $S_{\CG_k}$ the conformal WZW model with the current algebra
symmetry.  This theory possesses left and right-moving currents
$J^a (z)$ and $\Jb^a (\zb)$ satisfying the operator product
expansion (OPE) 
\beq
\label{2.1}
J^a (z) J^b (0) =  \frac{k}{z^2} \eta^{ab} + \inv{z}
f^{ab}_c \> J^c (0) + ....
\eeq
and similarly for $\Jb$, with $a=1,..{\rm dim} (\CG)$\cite{KZ}.   
For ordinary (bosonic) algebras we take $\eta^{ab} = \delta^{ab}/2$ 
corresponding to a normalization in the defining representation
of $tr (t^a t^b) = \delta^{ab}/2$.  For realizations based on free
fermions $J^a = \psi^\dagger t^a \psi$, the level $k=1$.   We will
need the Casimir in the adjoint representation defined by
$\eta_{ij} f^{jc}_k f^{ik}_d = C_{\rm adj} \delta^c_d$, where
$\eta_{ab}\eta^{bc} = \delta_a^c$.  

\def\alb{{\bar{\al}}}

We consider models with a number of flavors.  
Let $J^a_\al$, $\al = 1,2,..N_L$, and $\Jb^a_\alb$, 
$\alb = 1,2, ..N_R$, denote the resulting currents where
$\al, \alb$ are the flavor indices, $a$ are $\CG$ (color) indices,
and in general the number of left-moving flavors
 $N_L$ is not equal to $N_R$.  
The perturbations of the conformal field theory we will study
can be defined by the action
\beq
\label{2.2}
S = S_{\CG_1}^{N_L, N_R} + \int \frac{d^2x}{2\pi} ~ 
\sum_{\al, \alb, a}  g_{\al\alb} \> J^a_\al \Jb^a_\alb
\eeq
where $S_{\CG_1}^{N_L, N_R}$  is the formal action for
the WZW model at level 1 with $N_L, N_R$ numbers of chiral flavors. 
The above theory preserves the $\CG$ symmetry but introduces some
anisotropy in the flavor couplings $g_{\al\alb}$ which comprise
a $N_L \times N_R$ matrix.   

\def\gammab{\bar{\gamma}}
\def\Gb{\bar{G}}

The computation in \cite{GLM} is easily extended to this model. 
Operator product expansions involving the color indices 
give $C_{\rm adj}$ at each order, and one then only has to keep
track of the delta functions $\delta^{\al\beta}$ over the flavor
indices.  
Let 1A, 1B, 2A, 2B denote the 4 kinds of diagrams described in 
\cite{GLM}.  Define
\beq
\label{2.3}
G = \frac{k^2}{16} g g^T , ~~~~~~~~\bar{G} = \frac{k^2}{16} g^T g
\eeq
where $T$ denotes transpose. 
$G$ ($\bar{G}$) is a $N_L \times N_L$ ($N_R \times N_R$)  matrix.  
As such,  $G^n g $ is a $N_L \times N_R$ matrix, etc.  
The contributions to the \betaf function 
from the various diagrams at n-th order in
perturbation theory are
\barray
\beta^{(2A)}_{g_{\al\alb}} &=& \frac{C_{\rm adj}}{2} 
\( {G}^{(n-m-1)/2} g \)_{\al\alb} \( {G}^{(m-1)/2} \)_{\al\alb} 
\\
\label{2.4}
\beta^{(2B)}_{g_{\al\alb}} &=& \frac{C_{\rm adj}}{2} 
\( \frac{k}{4} \)^2 g_{\al\gammab} \({G}^{(m-3)/2} g \)_{\gamma\gammab}
\( {G}^{(n-m-1)/2} g \)_{\gamma\gammab} g_{\gamma\alb} 
\\
\beta^{(1A)}_{g_{\al\alb}} &=& - \frac{C_{\rm adj}}{2}
\( \frac{k}{4} \)^{-1} \(  ({G}^{(m-1)/2})_{\al\gamma} 
({G}^{(n-m)/2})_{\al\gamma} g_{\gamma\alb}  
+ g_{\al\gammab} ({\Gb}^{(n-m)/2} )_{\alb\gammab}
({\Gb}^{(m-1)/2} )_{\alb\gammab} \)  
\\
\beta^{(1B)}_{g_{\al\alb}} &=& - \frac{C_{\rm adj}}{2}
\( \frac{k}{4} \)^{-1} 
\( 
({G}^{(n-1)/2} )_{\al\al} g_{\al\alb} +  
({\Gb}^{(n-1)/2} )_{\alb\alb} g_{\al\alb} 
\)
\earray
(The $\gamma, \gammab$ indices are summed over.) 
The type 2A,B (1A,B) are for $n$ even (odd) order, and in all cases
$m$ is odd.  
For type 2A, $m=1,3,..,n-1$ whereas for 2B $m=3,5,..,n-1$. 
For type 1A, $m=3,5,..,n-2$.  

The series is easily summed with the following result.  Define
\beq
\label{2.5}
g' = \inv{1-G} g, ~~~~~ G' = \frac{G}{1-G}, ~~~~~\Gb' = \frac{\Gb}{1-\Gb} 
\eeq
Then 
\beq
\label{2.6} 
\beta_{g_{\al\alb}} = \frac{C_{\rm adj}}{2} 
\( 
(g'_{\al\alb})^2 + \frac{k^2}{16} g_{\al\gammab}(g'_{\gamma\gammab})^2 
g_{\gamma\alb} 
- \frac{4}{k} 
\( (G'_{\al\gamma} )^2 g_{\gamma\alb} + g_{\al\gammab} 
(\Gb'_{\alb\gammab})^2 
+ (G'_{\al\al} + \Gb'_{\alb\alb} )g_{\al\alb}  \) 
\) 
\eeq
(In the above formula $(g'_{\al\alb})^2$ is the square of the
matrix element $g'_{\al\alb}$ rather than the $\al\alb$ matrix
element of $g'^2$, and similarly for the $G$ terms. )

The resulting expressions for the \betaf function  are too complicated
to display here even in the case of 2 flavors. 
What is more important is that   
for the cases we have examined, 
the above \betaf function 
with $N_L= N_R$  has  no non-trivial fixed points at intermediate
values of the couplings.   We expect  then that when $N_L = N_R$ 
the only fixed points are the trivial ones at
$g=0$ or $\infty$  
as for the one-flavor case.

\section{Chiral Flavor Anisotropy} 

Consider the simplest case with unequal numbers of flavors, 
 $N_L = 1, N_R=2$, $k=1$, defined
by the action 
\beq
\label{3a}
S= S_{\CG_1}^{N_L=1, N_R =2}  + 
\int \frac{d^2x}{2\pi} \( g_1 J^a_1 \Jb^a_1 + g_2 J^a_1 \Jb^a_2 \) 
\eeq
The result eq. (\ref{2.6}) reduces to  
\barray
\beta_{g_1} &=&  C_{\rm adj}  \frac{ 
2  g_1 (g_1-4) \( 4g_1(g_1-4) - g_2^2 (g_2-g_1 - 4) \) }
{(g_1^2 + g_2^2 - 16)^2 } 
\nonumber
\\
\beta_{g_2} &=& \beta_{g_1} ( g_1 \to g_2, g_2 \to g_1 ) 
\label{3b} 
\earray

The fixed points are $g_1=g_2=g$ with $g=g_c = 2,4$.   
We computed the \betaf function for one other example with
$N_L =1, N_R=3$.  The only fixed points for all positive $g$
here are $g_{11}=g_{12} = 4/k, 2/k$ with $g_{13} = 0$, which
is equivalent to the previous example with $N_R=2$, and 
$g_{11}=g_{12}=g_{13} = 4/k, 4/3k$.  Based on these
examples, we expect that one generally flows to the isotropic
flavor manifold with $g_{\al\alb} = g$ for all $\al,\alb$ or
for a decoupled subset.  

When $g_{\al\alb} = g$ for all $\al,\alb$, the theory can be
reformulated as 
\beq
\label{3.1} 
S = S_{\CG_{k_L} \otimes \CG_{k_R} } 
+ \int \frac{d^2x}{2\pi} ~ g \> J^a \Jb^a 
\eeq
where the left-moving currents 
$J^a = \sum_\al^{N_L}  J^a_\al $  		
have level $k_L = N_L k$ and the right moving currents have level 
$k_R = N_R k$, and $S_{\CG_{k_L} \otimes \CG_{k_R} }$
is the formal action for the WZW model at these levels. 
 The \betaf function then reduces to 
\beq
\label{3.2} 
\beta_g = \frac{C_{\rm adj}}{2} 
\frac{ g^2 (1-k_L g/4)(1-k_R g/4) }{(1-k_L k_R g^2 /16 )^2 } 
\eeq
This \betaf function has two fixed points at $g=4/k_L$ and $4/k_R$. 
It also has two poles at $g=\pm 4/\sqrt{k_L k_R}$, however 
RG flows encounter the zeros before reaching these poles. 
Thus the fixed point exists within the perturbative domain.  
When $k_L = k_R$ the only fixed points are  at $g=0, \infty$.  
If $k_R > k_L$ and $C_{\rm adj} > 0$ then the RG flows from 
small coupling to $g_c = 4/k_R$.  

At the IR fixed point we expect on general grounds that  
\beq
\label{3.3} 
\beta_g = (2-\Gamma_{IR} ) (g-g_c ) + ....
\eeq
where 
$\Gamma_{IR}$ is the dimension of the irrelevant operator that
the perturbation $g J\Jb$ flows to in the IR.  We find  
\beq
\label{3.4} 
\Gamma_{IR} = 2 - \d_g \beta_g (g_c) =  2 \(1 + 
\frac{C_{\rm adj}}{k_R-k_L}
\) 
\eeq 

In ref. \cite{Andrei} thermodynamic Bethe ansatz analysis and other
more general arguments based on the equality of $c_R - c_L$ and
$k_R - k_L$ in the UV and IR indicate that the flow is to the
chirally asymmetric coset: 
\beq
\label{3.5} 
G_{k_L} \otimes G_{k_R} \longrightarrow 
\( \frac{G_{k_L} \otimes G_{k_R - k_L} }{G_{k_R}} \)_L 
\otimes \( G_{k_R - k_L} \)_R 
\eeq
The flow arrives at the IR fixed point via the primary operator
$\phi^{\rm adj}$ of the level $k= k_R - k_L$ current algebra   
with dimension 
$\Gamma_{IR} = 2 ( 1 + C_{\rm adj}/(k_R - k_L + \tilde{h}) )$
where $\tilde{h}$ is the dual Coxeter number of $\CG$. 
 ($\tilde{h} = N$ 
for $su(N)$.)   Note that when $k_L , k_R \to \infty$ but
$k_R - k_L $ is finite one recovers the results in \cite{PW}.   
Comparing with (\ref{3.4}) we see that our \betaf function finds this
fixed point except  
for the shift of $k$ by the dual Coxeter number.  
The origin of this discrepancy is unclear, and for the remainder
of this section   
we only discuss the possible origins of it.   

The numerous tests of the \betaf function performed in \cite{BLdual} 
strongly indicate  that the result is correct for when  
$k_L= k_R=1$.  
In particular, the case of imaginary potential sine-Gordon theory
has a fixed point both in the ultra-violet and infra-red, and the
anomalous dimensions at both points as computed from the \betaf function
agree with exact Bethe ansatz calculations.  
The dependence of $\Gamma_{IR}$ on the difference    $k_R-k_L$
relies on all the higher order contributions to the \betaf function;
keeping only the two-loop contribution would give
$\Gamma_{IR} = 2 (1 + C_{\rm adj}/(k_R + k_L) )$.  This suggests 
that the \betaf function in \cite{GLM} is not just the leading
terms in $1/k_L, 1/k_R$.

The shift by the dual Coxeter number 
is the same shift that appears in the affine-Sugawara construction:
$T =  J^a J^a/(k+\tilde{h})$  
which arises from proper normal-ordering\cite{Olive} of the product
of two currents.  So one possibility is that there are additional
corrections to the \betaf function coming from the normal-ordering
of $J\Jbar$.   However this seems unlikely since such corrections
would appear at level 1.    
The spectrum of primary fields in the current algebra depends on 
the level $k$\cite{KZ}.   For example for 
$su(2)$  at level $k=1$ the field 
$\phi^{\rm adj}$ is not even  in the spectrum and the correct
result is   
$\Gamma_{IR} = 4$ corresponding 
to a perturbation $T \bar{T}$ where $T$ is the stress tensor. 
Since $\Gamma_{IR}$  is not a universal function but rather 
takes different forms depending on $k$ and reflecting the fusion
rules, 
for the \betaf function to yield this  result 
it would have to know about the spectrum of primary fields and would
have to possess several zeros depending on the value of $k$\footnote{This
is actually reminiscent  of the dependence of the 
phase structure of Yang-Mills theory on the numbers of chiral flavors 
and colors.  See e.g. \cite{wilczek}\cite{seiberg}. }.   
.   
In \cite{GLM} the \betaf function was computed using the formula
\beq
\label{iso}
\langle X \rangle =  F(g, \log a) \int \frac{d^2x}{2\pi} 
\langle J\Jbar (x) X \rangle
\eeq
where $F$ is a function 
of the  ultra-violet cutoff $a$  and $X$ is an arbitrary field 
or product of fields.   The divergences found in \cite{GLM} 
are independent of which field $X$ appears in the above equation. 
Thus another possibility is that there are additional wavefunction
renormalizations  
that depend on what the field $X$ is.  This resolution would have
the opportunity to depend intricately on the level $k$ since
the spectrum of $X$ depends on $k$.

\section{Color and flavor anisotropy}

\def\ab{{\bar{a}}}

\def\Ctilde{{\tilde{C}}}

The most general model with flavor and color anisotropy 
corresponds to a perturbation $\sum_A g^A_{\al\alb} d^A_{ab} 
J^a_\alpha \bar{J}^b _\alb $ where $g^A_{\al\alb}$ are
couplings and  $d^A_{ab}$ are fixed 
quadratic forms that characterize the color anitotropy.
   It is possible to find
the \betaf function for this case however the formulas are rather
cumbersome.  In this section we limit ourselves to 
$g^A_{\al\alb} = g_A$ for all $\al, \alb$.   As in the previous
section the theory is then defined by 
\beq
\label{4.1} 
S = S_{G_{k_L} \otimes G_{k_R}} + \int \frac{d^2x}{2\pi} 
\sum_A g_A ~ d^{A}_{ab} \> J^a  \bar{J}^b  
\eeq
where $J^a$ ($\Jb^\ab$) has level $k_L = N_L k$ ($k_R = N_R k$). 
The \betaf function in \cite{GLM} is easily generalized to the
case $k_L \neq k_R$ by keeping track of the $k$ terms in 
$JJ$ verses $\Jb \Jb$ OPE's.   As in \cite{GLM}\cite{Lec}
 define the RG
data $C, \Ctilde, D$  in terms of the OPE's 
\beq
\label{4.2}
\CO^A (z, \zbar) \CO^B (0) \sim \inv{z\zbar}
\sum_C C^{AB}_C \> \CO^C (0)
\eeq
\beq
\label{4.3}
T^A (z) \CO^B (0) \sim \inv{z^2}
\( 2k_L  D^{AB}_C +  \Ctilde^{AB}_C  \) \CO^C (0)
\eeq
where $T^A = d^A_{ab} J^a J^b$.   Let $D(g)$ be the matrix of
couplings 
\beq 
\label{4.4} 
D(g)^A_B = \frac{\sqrt{k_L k_R}}{2} \sum_C g_C D^{AC}_B 
\eeq
and let 
\beq
\label{4.5}
g' = g (1-D^2 (g))^{-1} 
\eeq
Then the result is  
\beq
\label{4.6}
\beta_g = -\inv{2} C(g', g') (1+D^2) 
+  \frac{k_L + k_R}{2\sqrt{k_L k_R}}  
\( C(g'D, g'D) D - \Ctilde (g'D, g) \) 
\eeq

\def\ct{\Ctilde}

As an example we work out the $su(2)$ case.   Let us normalize
the currents as 
\beq
\label{4.7}
J_3(z) J_3(0) \sim \frac{k_L}{2} \inv{z^2} , ~~~~~~~
J_3(z) J^\pm (0) \sim \pm \inv{z} J^\pm (0)
,~~~~~
J^+ (z) J^-(0) \sim \frac{k_L}{2} \inv{z^2} + \inv{z} J_3(0)
\eeq
and consider the action
\beq
\label{4.8}
S  =  S_{su(2)_{k_L} \otimes su(2)_{k_R} }  +  
\int \dx \(  g_1 (J^+ \Jb^- + J^- \Jb^+ )
+ g_2 J_3 \Jb_3  \)
\eeq
The RG data was computed in \cite{GLM}:
$C^{12}_1 = C^{21}_1 = -1$, $C^{11}_2 = -2$,
$\ct^{11}_1 = \ct^{21}_1 = 1$,  $\ct^{12}_2 = 2$,
$D^{11}_1 = D^{22}_2 = 1/2$. 
The result is 
\barray
\nonumber
\beta_{g_1} &=& \frac{ 
16 g_1 \( k_L k_R g^2_1 g_2 + 16g_2 - 2 (k_L + k_R) (g_1^2 + g_2^2 ) \) }
{(k_L k_R g_2^2 - 16)(k_L k_R g_1^2 - 16 ) }
\\
\label{4.9} 
\beta_{g_2} &=& \frac{ 
16 g_1^2 (k_R g_2 - 4)(k_L g_2 - 4) }
{ (k_L k_R g_1^2 - 16 )^2 } 
\earray

The zeros  of the above \betaf function are 
(i)  $g_1 = g_2=4/k_R, 4/k_L$ and 
(ii) $g_1 = -g_2 = 4/k_R , 4/k_L$.   
Thus one can flow to the $su(2)$ color isotropic manifolds 
$g_1 = \pm g_2$ and then to the fixed points of the kind described
in section III.  When $k_R = k_L$ there are regions that are  
attracted to the isotropic manifolds, but then flow off to 
$\infty$\cite{BLdual} away from the isotropic line.   
Here, one instead reaches the fixed point
before reaching the pole, and the flow is again chirally stabilized.

\section{Discussion}

When one of the levels $k_L$ or $k_R$ 
equals zero,  the model is related to the important problem of
the Kondo lattice\cite{Andrei2}.   Thus some of the models we are 
considering here can be viewed as intermediate between bulk 
and boundary perturbations.   It would be interesting 
to generalize our computation of the $\beta$ function to the case of  
purely boundary interactions, such as in the Kondo model. 
This could have some applications to the problem of tachyon condensation
in string theory\cite{string}.

\section{Acknowledgments}

I wish to thank N. Andrei, D. Bernard and F. Smirnov  for discussions.  
This work is in part supported by the NSF.

\end{document}